\documentclass{article}
\PassOptionsToPackage{preprint}{neurips_2025}
\usepackage{neurips_2025}

\usepackage[utf8]{inputenc}
\usepackage[T1]{fontenc}
\usepackage{hyperref}
\usepackage{url}
\usepackage{booktabs}
\usepackage{amsfonts}
\usepackage{amsmath,amssymb}
\usepackage{nicefrac}
\usepackage{microtype}
\usepackage[dvipsnames,table]{xcolor}

\usepackage[pdftex]{graphicx}
\usepackage{placeins}   
\usepackage{enumitem}

\title{TalkingMachines: Real-Time Audio-Driven FaceTime-Style Video via Autoregressive Diffusion Models}

\author{%
  Chetwin Low\thanks{Equal contribution.}%
  \quad Weimin Wang\footnotemark[1]\\
  Character AI\\
  \texttt{\{chetwinlow, weiminwang\}@character.ai}
}

\begin{document}
\maketitle

\begin{abstract}
In this paper, we present TalkingMachines—an efficient framework that transforms
pretrained video generation models into real-time, audio-driven character animators.
TalkingMachines enables natural conversational experiences by integrating an
audio large language model (LLM) with our video generation foundation model.
Our primary contributions include:
(1) We adapt a pretrained SOTA image-to-video DiT into an audio-driven avatar
generation model of 18 billion parameters;
(2) We enable infinite video streaming without error accumulation through
asymmetric knowledge distillation from a bidirectional teacher model into a sparse
causal, autoregressive student model;
(3) We design a high-throughput, low-latency inference pipeline incorporating several
key engineering optimizations such as:
~(a) disaggregation of the DiT and VAE decoder across separate devices,
~(b) efficient overlap of inter-device communication and computation using CUDA streams,
~(c) elimination of redundant recomputations to maximize frame-generation throughput. Please see demo videos here - https://aaxwaz.github.io/TalkingMachines/

\end{abstract}

\FloatBarrier

\vspace{0.5cm}
\begin{figure}[!htb]
  \centering
  \includegraphics[width=0.9\textwidth]{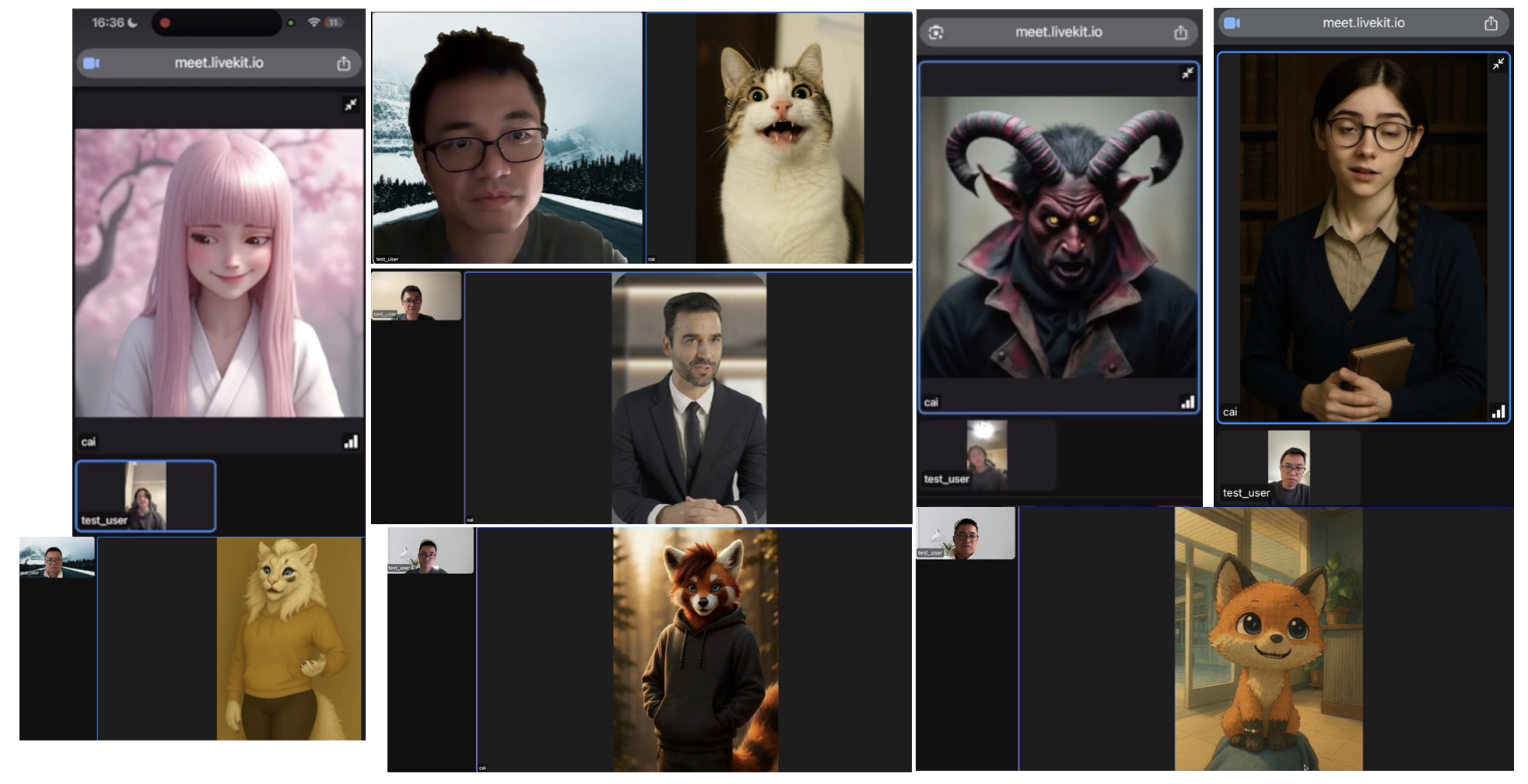}
  \caption{TalkingMachines provides a framework to generate highly dynamic, immersive
    FaceTime experiences based on different character styles.}
  \label{fig:teaser}
\end{figure}
\vspace{0.3cm}

\section{Introduction}

The emergence of video foundation models has revolutionized content creation, enabling high-quality video generation from text prompts and images. However, existing models face significant challenges when applied to real-time interactive applications like video conferencing. The primary limitation is their reliance on bidirectional attention mechanisms, which require processing the entire video sequence all at once—where each individual frame relies on future frames. This architectural constraint introduces prohibitive latency for streaming applications.

In this work, we present TalkingMachines, a novel approach that transforms a pretrained bidirectional video diffusion model into an efficient autoregressive system capable of real-time, audio-driven character animation. Our method addresses three key challenges:

\begin{enumerate}[leftmargin=*,noitemsep,topsep=0pt]
\item \textbf{Adaptation of pretrained models:} We leverage the knowledge embedded in large-scale video foundation models while adapting them for audio-driven generation through specialized attention mechanisms and training strategies.
\item \textbf{Infinite-length generation:} Through our modified Distribution Matching Distillation (DMD)~\cite{yin2024one,yin2024improved,yin2024slow} approach, we enable autoregressive generation without the error accumulation typically associated with sequential predictions, allowing for arbitrarily long video streams.
\item \textbf{Real-time performance:} By distilling the model to 2 diffusion steps and implementing system-level optimizations, we achieve real-time latency suitable for interactive applications.
\end{enumerate}

\section{Background}

\subsection{Flow Matching Models}

In our framework, we leverage a pretrained image-to-video (I2V) model from WAN 2.1 \cite{wan2025}, which was originally trained using the flow matching paradigm. We adapt this model for our TIA2V (Text-Image-Audio to Video) generation task by introducing audio as an additional conditioning signal alongside the existing image and text inputs.

Flow matching provides a stable and efficient alternative to traditional stochastic diffusion training by modeling deterministic trajectories in latent space via velocity prediction. Rather than sampling with stochastic differential equations, flow matching treats generation as an ordinary differential equation (ODE) integration problem, enabling more stable training and inference.

During fine-tuning, given a target latent $x_1$ (i.e., a video latent from ground truth frames), we sample a Gaussian noise latent $x_0 \sim \mathcal{N}(0, I)$ and a timestep $t \in [0, 1]$ from a logit-normal distribution. We then compute an intermediate latent using a linear interpolation, following the Rectified Flow formulation \cite{esser2024rectified}:

\begin{equation}
x_t = t x_1 + (1 - t) x_0
\end{equation}

The ground truth velocity is simply the difference between target and noise latents:
\begin{equation}
v_t = \frac{dx_t}{dt} = x_1 - x_0
\end{equation}

We condition the model on three modalities: text embedding $c_{txt}$, an initial reference image $c_{img}$, and audio embeddings $c_{aud}$. The model is trained to predict the velocity field from this conditioning. The training objective becomes:

\begin{equation}
\mathcal{L}_{FM} = \mathbb{E}_{x_0, x_1, t} \left[ \left\| u(x_t, c_{txt}, c_{img}, c_{aud}, t; \theta) - (x_1 - x_0) \right\|^2 \right]
\end{equation}

where $u(x_t, c_{txt}, c_{img}, c_{aud}, t; \theta)$ denotes the predicted velocity from the model. This adaptation retains the benefits of flow matching—such as training stability and efficient sampling—while extending the model’s expressiveness to multimodal conditioning. By incorporating temporally aligned audio features into the pretrained I2V architecture, we enable high-quality, lip-synced video generation driven by real-time conversational audio inputs.

\subsection{Distribution Matching Distillation}
Distribution Matching Distillation (DMD) addresses the computational burden of multi-step sampling in diffusion models by training efficient few-step generators. Unlike traditional distillation methods that preserve individual sampling trajectories, DMD operates at the distribution level, enabling greater flexibility in student model design.

The fundamental principle behind DMD involves minimizing the reverse KL divergence between distributions generated by a multi-step teacher model and a few-step student generator. The optimization objective can be expressed through the reverse KL divergence:
\begin{equation}
\mathcal{L}_{\text{DMD}} = \mathbb{E}_t \left[ \text{KL}(p_{\text{gen},t} \| p_{\text{data},t}) \right]
\end{equation}

The gradient of this objective admits a tractable approximation:
\begin{equation}
\nabla_\phi \mathcal{L}_{\text{DMD}} \triangleq \mathbb{E}_t \left[ \nabla_\phi \text{KL}(p_{\text{gen},t} \| p_{\text{data},t}) \right]
\end{equation}

\begin{equation}
\approx -\mathbb{E}_t \left[ \int \left( s_{\text{data}}(\Psi(G_\phi(\epsilon), t), t) - s_{\text{gen}}(\Psi(G_\phi(\epsilon), t), t) \right) \frac{\partial G_\phi(\epsilon)}{\partial \phi} d\epsilon \right]
\end{equation}

where $G_\phi$ denotes the student generator parameterized by $\phi$, $\Psi$ represents the forward diffusion process, and $s_{\text{data}}, s_{\text{gen}}$ are score functions estimated from the data and generated distributions respectively.
\section{Related Work}

\subsection{Video Foundation Models}

Recent advances in video generation have been driven by diffusion-based approaches, particularly those using Transformer architectures. State-of-the-art models~\cite{videoworldsimulators2024, hong2022cogvideo, wan2025, kong2024hunyuanvideo, genmo2024mochi} have demonstrated impressive generation quality and visual fidelity. However, these models are primarily designed for offline generation, where the entire video sequence is processed at once. The bidirectional attention mechanisms employed in these architectures, while beneficial for maintaining temporal coherence and visual quality, introduce prohibitive latency for real-time streaming applications where frames must be generated on-demand.

\subsection{Audio-Driven Animation}

Previous work on audio-driven facial animation has explored various approaches, from 3D morphable models to GAN-based methods. Recent diffusion-based approaches like EMO\cite{tian2024emo} and Hallo3\cite{cui2024hallo3} have shown promise but still suffer from high latency and limited style diversity. Our work builds on these foundations while addressing their limitations for real-time applications.

\subsection{Real-time Talking Head Generation}
Recent real-time talking head generation methods, such as VASA-1~\cite{xu2024vasa1lifelikeaudiodriventalking} and MegaPortraits~\cite{drobyshev2023megaportraitsoneshotmegapixelneural}, employ a common architecture centered around a disentangled facial latent space. These approaches typically use an appearance encoder to extract static identity features, and a motion encoder or predictor to represent facial dynamics as a low-dimensional temporal sequence—often compact 1D vectors. At inference time, a lightweight generative model predicts the motion sequence from audio and expression cues, which is then combined with the static appearance embedding via a decoder to synthesize realistic talking head animations in real time.

The computational efficiency of these methods enables real-time generation on consumer-grade GPUs without requiring advanced optimization techniques. However, their architectural assumptions inherently limit generality. Specifically, the reliance on rigid and non-rigid 3D warping tailored for faces makes these models difficult to extend to full-body avatars, non-realistic videos such as animations or general-purpose video synthesis. Additionally, the compactness of the motion representation, while efficient, constrains the model's ability to capture complex or large-scale motion beyond head and facial dynamics.

In contrast, our approach explores the use of large-scale Diffusion Transformers, which offer strong generative capabilities. To meet the demands of real-time performance, we complement these models with sparse causal attention and targeted engineering optimizations, enabling synthesis beyond facial domains while preserving responsiveness.

\subsection{Model Distillation}

Knowledge distillation has emerged as a powerful technique for accelerating diffusion models. Progressive distillation and consistency models have shown success in reducing sampling steps. DMD offers unique advantages by allowing architectural flexibility between teacher and student models, which we exploit to enable the bidirectional-to-causal transformation.

\section{Method}

\subsection{Model Architecture}

Our model architecture builds upon the WAN2.1 \cite{wan2025} 14B I2V model, which supports video generation from a ground truth image and text prompt. We extend this with:

\textbf{Audio Cross Attention with Attention Mask:} We modified the Cross Attention layer to include support for the input of audio sequence. Basically within each Cross Attention layer, we initialize new layers to generate keys and values from the input audio tokens, which are then cross-attended to the queries of the corresponding latent frame tokens. Similar to MagicInfinite \cite{yi2025magicinfinite}, we employ local attention masks that focus on facial regions during audio cross-attention. This improves lip-sync accuracy by reducing interference from non-facial tokens such as noise and background visual signals.

\textbf{Audio Projection Layer:} A 1.2B parameters audio module that processes raw audio token embeddings before passing into cross-attention layers. Inspired by Hallo3\cite{cui2024hallo3}, the audio tokens are temporally aligned with video frames using a local window-based approach: each latent frame will attend only to the audio tokens within a window of five latent frames centered around itself. Considering that speech-driven motion—such as lip movement, head gestures, and body dynamics—is predominantly influenced by temporally local audio features, this localized attention helps the model to learn audio-visual correlations faster and better.

\textbf{Speaking/Silence Mode:} For each face of the training clip, we detect whether it is actually in \textit{Speaking} or \textit{Silence} mode. When it is in \textit{Silence} mode, we will use a special embedding (e.g. zero embedding) instead of the original audio embedding to represent the status, and pass that for cross attention. We are able to use a lipsyncing evaluation model such as SyncNet\cite{chung2017out} to detect the (face, audio) similarity score (i.e. conf). For videos that have multiple faces, we will select the face that has the highest SyncNet score as the face in \textit{Speaking} mode, and the rest in \textit{Silence} mode. We also manually filter the training data in \textit{Silence} mode to ensure there are no dubbed cases. During inference, we will enable the \textit{Silence} mode for the given character by passing in the corresponding special embedding when it is the character's listening phase. This also makes it possible for Multi Character generation cases - where there are multiple faces within the video being generated, and we are able to switch the modes among them depending on the turns of speech.
\subsection{Asymmetric Distillation with Modified CausVid}

The core innovation of our approach is extending and modifying the CausVid \cite{yin2024slow} framework, which distills the knowledge of a bidirectional teacher model into an autoregressive student model. Similar to CausVid, to reduce first frame latency, we break up the original context length of our bidirectional teacher (21 latent frames) into smaller chunks, maintaining bidirectional attention within chunks, but introducing causal attention masking across chunks, enabling the autoregressive student to generate shorter segments of video frames sequentially.

However, our approach diverges from CausVid in several key aspects. CausVid employs an approach akin to Diffusion Forcing \cite{chen2024diffusion}, where each chunk is assigned an independent randomly sampled timestep of gaussian noise. As a result, chunks attend to preceding chunks that may be at different timesteps of noise. In contrast, we observe that few-step generators typically exhibit a piecewise flow trajectory. Allowing chunks to attend to other chunks of mismatched timesteps would be suboptimal. To address this, we maintain the sampled timestep as a global attribute shared across all chunks within a video, aligning more closely with the pre-training set up. Furthermore, we observe that speech-driven motion only requires a small temporal attention window to have smooth temporary continuity, as such we designed the following sparse causal attention masking, effectively reducing inference memory requirements by requiring a smaller KV cache and also faster inference due to small attention KV context length, without any quality degradation.

For each 81-frame video clip (21 latent frames after VAE encoding), we:
\begin{enumerate}
\item \textbf{Chunked Processing:} Divide the latent sequence into 7 chunks of 3 latent frames each
\item \textbf{Sparse Attention Pattern:} Each token in chunk $c_t$ attends to:
\begin{itemize}
\item All tokens within the current chunk $c_t$
\item All tokens in the previous chunk $c_{t-1}$
\item All tokens in the starting chunk $c_0$ (containing the ground truth image)
\end{itemize}
\end{enumerate}

This attention pattern ensures temporal continuity while preventing error accumulation, as the model always has access to the clean reference image in $c_0$.

The mathematical formulation of our sparse attention can be expressed as:
\begin{equation}
\text{Attention}(Q_t, K, V) = \text{softmax}\left(\frac{Q_t K^T}{\sqrt{d}}\right)V
\end{equation}
where $K$ and $V$ include tokens from $\{c_0, c_{t-1}, c_t\}$ only.

\begin{figure}[h]
\centering
\includegraphics[width=0.9\textwidth]{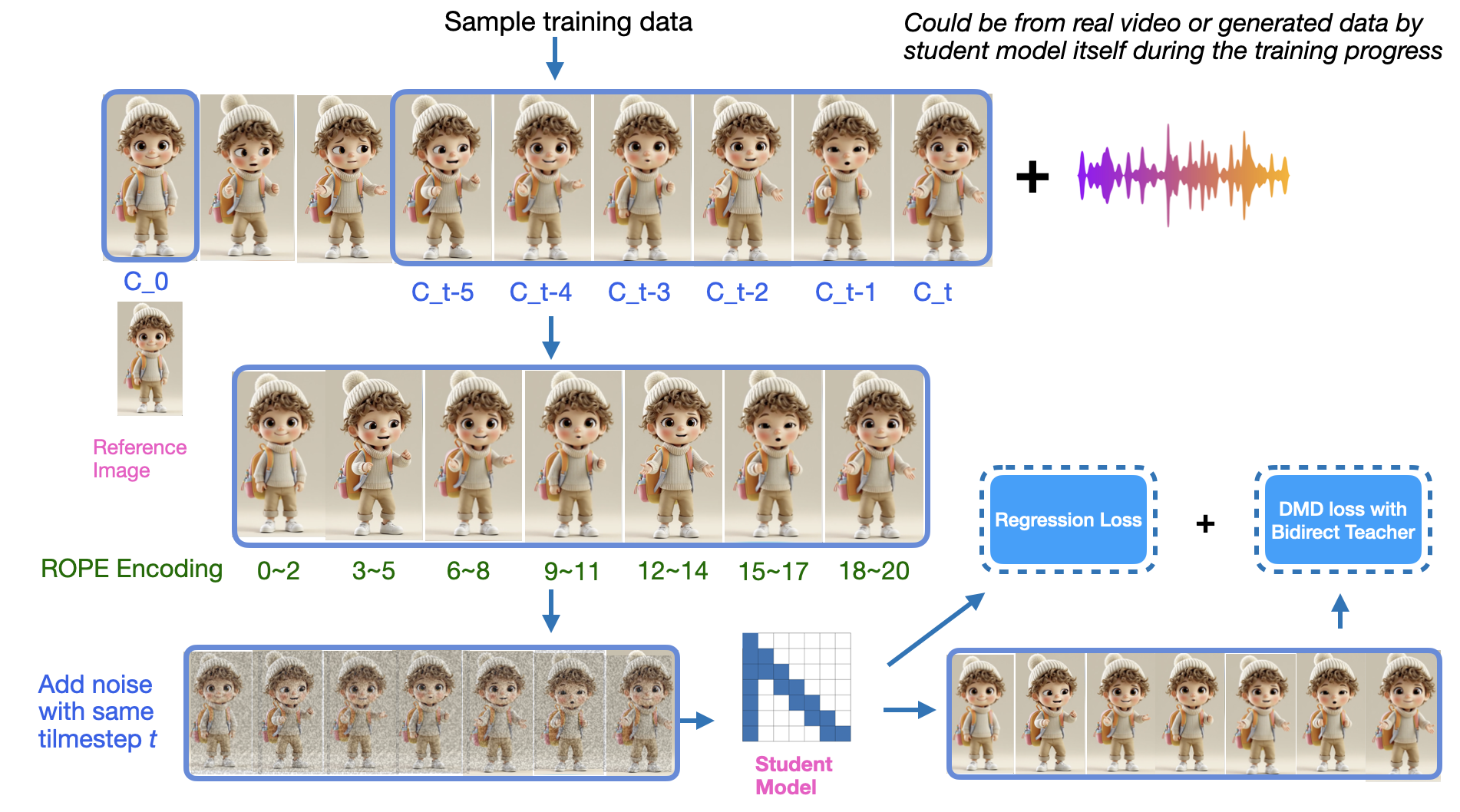}
\caption{Overview of the DMD training workflow for TalkingMachines. The diagram illustrates the asymmetric distillation process where a bidirectional teacher model is distilled into an autoregressive student model. The training incorporates mixed data generation with synthetic samples from the student model, sparse causal attention patterns across chunks, and a combination of DMD loss with regression loss for stable convergence.}
\label{fig:dmd_workflow}
\end{figure}

We employ DMD to distill our model from 24 NFEs (12×2 with CFG) to just 2 NFEs. The key modifications include the following (please refer to Figure~\ref{fig:dmd_workflow} for illustrated workflow).
\begin{enumerate}
\item \textbf{Mixed Training Data:} We combine real training clips with synthetic samples generated by the student model using a (image, audio) pair in the sparse causal attention, improving generalization to unseen styles. We do this in progressive training fashion - we start with only real training data until the student model converges to some extent, before we leverage it to generate synthetic samples on the fly during training. 
\item \textbf{Sparse Causal Attention:} Each chunk only attends to current chunk $c_t$ (bidirectional within current chunk), previous chunk $c_{t-1}$, and the starting chunk $c_0$.
\item \textbf{Regression Loss without GAN:} We added regression loss on top of student model's prediction in addition to DMD loss to stabilize the training.
\end{enumerate}

\subsection{System Optimizations}

\textbf{Score-VAE Disaggregation:} Traditionally, latent diffusion models (LDMs)\cite{rombach2022highresolutionimagesynthesislatent} exhibit significantly higher computational latency during the denoising process compared to the subsequent VAE \cite{kingma2019introduction} decoding stage—often to the extent that decoding latency is considered negligible. However, with the growing adoption of transformer-based\cite{vaswani2017attention} architectures as the standard backbone for diffusion models, this assumption no longer holds. The performance gap between the diffusion model and the VAE is narrowing, owing to the proliferation of hardware- and software-level optimizations for transformers, including fast attention kernels\cite{dao2022flashattention, dao2023flashattention2} and fused MLP operations. Furthermore, in the context of real-time generation, where latency constraints are stringent, any inefficiency is non-trivial, motivating our Score-VAE disaggregation server design to support real-time generation.

In a single-GPU configuration, VAE decoding and device-to-host (d2h) transfer of the output tensor account for approximately 30\% of the end-to-end generation time per video chunk. This bottleneck is further amplified when employing Sequence Parallelism (SP)\cite{jacobs2023deepspeed} to distribute the diffusion computation across multiple GPUs. As the diffusion workload becomes increasingly parallelized, the relative cost of VAE decoding dominates, ultimately limiting scalability and becoming the primary obstacle to real-time streaming performance.

\begin{figure}[h]
\centering
\includegraphics[width=0.8\textwidth]{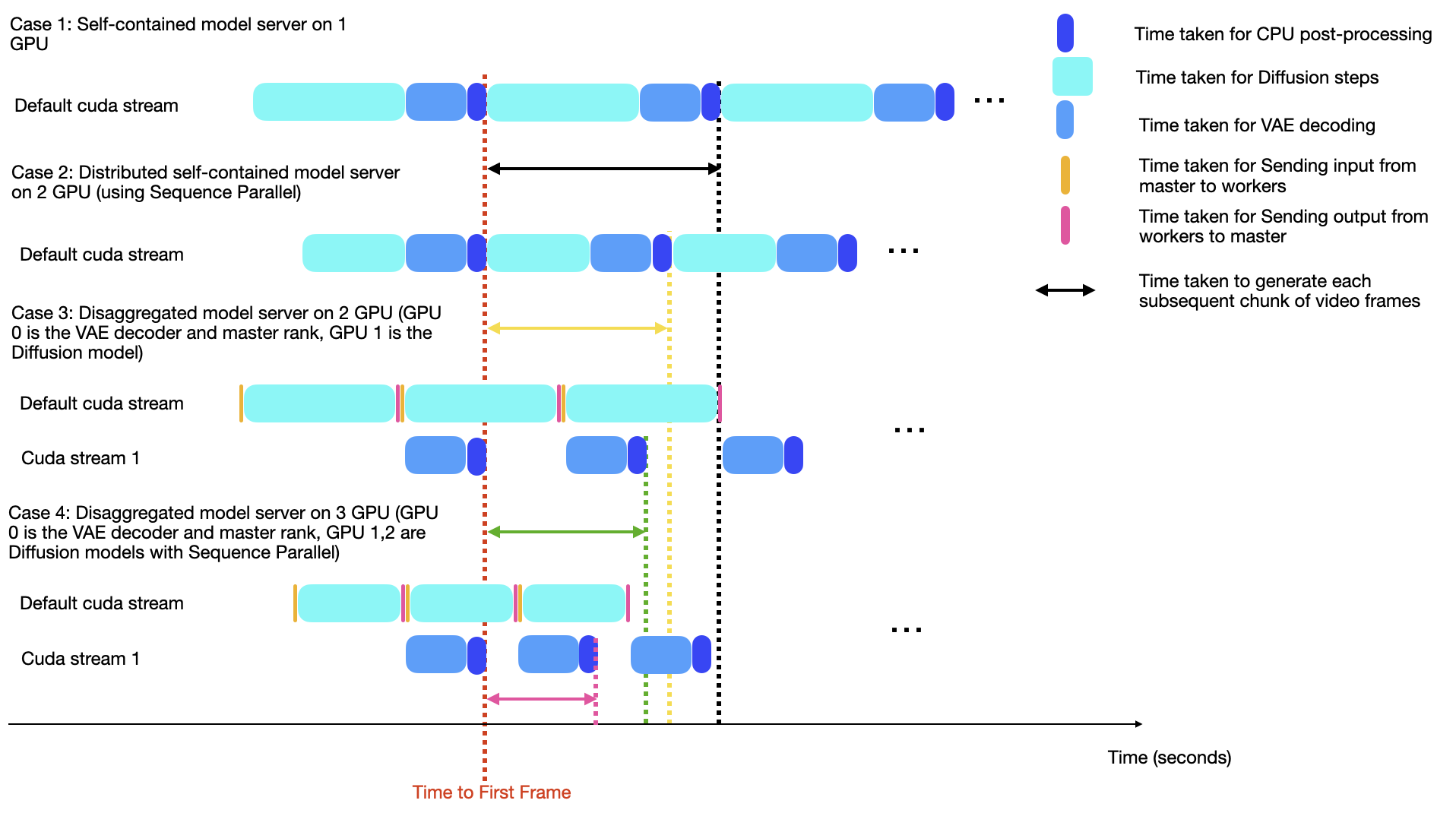}
\caption{Runtime analysis comparing the latency of various server designs like a simple self-contained server, and our Score-VAE disaggregation server, both with and without Sequence Parallelism.}
\label{fig:runtime}
\end{figure}

To evaluate whether a system meets real-time requirements, we define a key success criterion: the time taken to generate a sequence of video frames must be consistently less than the duration of the video sequence itself. To measure this, we introduce the Time Taken Between Chunks (TTBC) as a metric for comparison across different server configurations (illustrated in Figure~\ref{fig:runtime}). In Cases 1 and 2, where self-contained model servers are used (i.e., each GPU independently performs both diffusion and decoding), we observe diminishing returns when scaling from 1 to 2 GPUs. The improvement in TTBC is minimal, and neither configuration achieves the latency required for real-time generation. In contrast, Case 3 implements a 2-GPU disaggregated server, where one GPU is dedicated to diffusion (worker) and the other to VAE decoding (master). This design allows the worker to maximize diffusion throughput while overlapping decoding operations on the master GPU. Even without distributed diffusion, the 2-GPU disaggregated server outperforms the self-contained server with same number of GPUs in TTBC. Finally, in Case 4, we scale the number of diffusion workers to two while maintaining a single dedicated decoder. This configuration demonstrates superior scalability, with improved TTBC compared to Case 2. Both Cases 3 and 4 are able to consistently meet the real-time threshold for video generation.

\textbf{Efficient computation-communication overlap with CUDA streams:} The disaggregation server design introduces additional inter-device communication, as the master rank must broadcast the inputs to, and receive outputs from, the workers ranks. We leverage NCCL collectives to manage these data transfers efficiently. Crucially, we make use of additional CUDA\cite{nickolls2008scalable} streams to overcome the blocking nature of these NCCL collectives and ensuring the VAE decoding can be done in parallel with these collectives, thereby significantly improving inference throughput.


\textbf{KV \& Embedding Cachings:} During inference, we cache all key-value pairs for chunk $c_{t-1}$ and $c_0$, for each timestep, over all transformer blocks, eliminating redundant computations. Other embeddings such as timestep embedding, guidance embedding, and context embedding are also cached to improve inference speed.

\section{Experiments}

\subsection{Datasets}

We prepare high quality, human centric videos containing 1.5M clips that are first split by scene detection \cite{Castellano}, and then are filtered based on aesthetics\cite{laion-aesthetics} and motion scoring\cite{teed2020raft}, as well as using OCR\cite{paddleocr2020} model to filter out videos that contain too much text. The clips are with minimum 4s in duration, and contain speakers of diverse poses and distances to camera. The majority of the clips are focused on a single speaker, with a small percentage containing multiple speakers.

Our model's target resolution is 512x512, and supports diverse aspect ratios of input video frames that fall within the total area of 512x512. Using a lower resolution than the original pretrained model's resolution of 480x832 also makes inference faster as it reduces the context length of the transformer blocks, which reduces the burden for streaming. Based on our internal evaluation, 512x512 works well for both desktop and mobile-based devices. 

\subsection{Training Setup}
\label{subsec:training setup}

Our end to end training is split into three stages. 1) WAN 2.1  pretrained model warm-up. 2) Audio pretraining. 3) Distillation and autoregressive training. Our training starts with WAN2.1 I2V 480P, which has the original resolution of 480x832, and larger than our target resolution of 512x512. 

\begin{itemize}
\item \textbf{Stage 1 Warm up:}
\begin{itemize}
\item Total training steps: 9k
\item GPUs: 128 H100 for 1 day
\item Target: adapting the pretrained model to our new resolution as well as human centric training data.
\item Model weights: Learning all layers without freezing
\end{itemize}
\item \textbf{Stage 2 Audio Pretraining:}
\begin{itemize}
\item Total training steps: 30k
\item GPUs: 384 H100 for 5 days
\item Target: Learning the new audio layers and lip-syncing capabilities. 
\item Model weights: Initializing new audio layers, and freezing all non-audio parameters. 
\end{itemize}
\item \textbf{Stage 3 Sparse Autoregressive Distillation:}
\begin{itemize}
\item Total training steps: 20k
\item GPUs: 128 H100 for 10 days
\item Target: Distilling the model into 2 diffusion steps, with causal sparse attention. 
\item Model weights: Learning all layers without freezing. 
\end{itemize}
\end{itemize}

\subsection{Training Infrastructure and Strategies}
As described in section\ref{subsec:training setup}, we choose 512x512 as the target resolution not only for real-time generation, but also faster training iterations. At this resolution, the DiT model parameters, gradients and optimizer states consume approximately 204GB of GPU memory in bfloat16, while activations reaching up to several terabytes of GPU memory. Through activation checkpointing and parameter sharding, we are able to fit the peak activation memory within each GPU, eliminating the need for model parallelism and enabling training with data-parallel variants alone. 

In addition to the DiT backbone, training requires several auxiliary models—namely, the VAE, and Text, Image, and Audio encoders—to produce the necessary embeddings. While each of these modules is smaller than DiT, they still contribute significant memory overhead. To optimize memory allocation for activations, we employ DeepSpeed ZeRO Stage 3~\cite{rajbhandari2020zeromemoryoptimizationstraining} to shard the encoder modules. This strategy frees over 20GB of GPU memory with negligible impact on per-step performance, striking a favorable trade-off between embedding generation time and activation capacity.

Stage 1 and Stage 2 training of the DiT model are conducted using DeepSpeed ZeRO Stage 2 on NVIDIA H100 GPUs, leveraging GPUDirect-TCPX for high-speed interconnect. This configuration provides a strong baseline for efficient scaling under data-parallel training.

In Stage 3, corresponding to the DMD training, the training setup requires three full copies of the diffusion model to reside in GPU memory—two of which are trainable—totalling 450GB of GPU memory, thereby necessitating a more aggressive memory management strategy. Initial attempts using ZeRO Stage 3 are found to scale poorly as GPU counts increase, primarily due to its global parameter sharding, which introduces substantial cross-device communication overhead.

To address this bottleneck and improve throughput, we implement Hybrid Sharded Data Parallel (HSDP) with PyTorch FSDP\cite{zhao2023pytorchfsdpexperiencesscaling} API. This strategy enables parameter, gradient, and optimizer states sharding within localized GPU groups, thereby significantly reducing cross-node communication and improving memory efficiency. 

Additionally, experiments are conducted on a cluster equipped with NVIDIA H200 GPUs and InfiniBand. We observe a 2x speed up in per-step training time, primarily due to faster inter-node bandwidth and the ability to fit all relevant tensors within a single node, effectively eliminating the overhead of cross-node sharding.

Scaling training to a 720×720 resolution increases the total activation memory footprint by approximately 3–4×, which exceeds the capacity of a single GPU even with activation checkpointing enabled. Since both ZeRO and FSDP are fundamentally data-parallel approaches, no amount of sharding will reduce activations. 

To address this, we rely on Sequence Parallelism (SP)\cite{jacobs2023deepspeed}, a form of intra-layer parallelism that partition sequences across devices, hence reducing per-GPU computation. In our implementation, we integrate SP as part of a 2D parallelism strategy, where we shard model parameters, gradients and optimizer states (via ZeRO/HSDP), while splitting the computation and activations further within a smaller set of GPUs.

\subsection{Distillation Ablations}

To better understand the trade-offs between chunk size and diffusion steps in our distillation process, we conduct an ablation study using a 2×2 factorial design. Specifically, we vary the chunk size (3 vs. 7) and the number of diffusion steps (2 vs. 4), and evaluate performance using perceptual metrics and GPU efficiency, based on an internal video dataset containing mainly realistic videos of talking persons. The metrics include Fréchet Video Distance (FVD) \cite{unterthiner2019fvd} and two variants of SyncNet-based lip-sync confidence: Sync-C and Sync-D \cite{chung2017out}.

\begin{table}[h]
\centering
\begin{tabular}{ccccccc}
\toprule
\textbf{Chunk Size} & \textbf{Steps} & \textbf{FVD $\downarrow$} & \textbf{Sync-C $\uparrow$} & \textbf{Sync-D $\downarrow$} & \textbf{Score GPUs} & \textbf{VAE GPUs} \\
\midrule
7 & 4 & 208.2 & 6.28 & 8.94 & 4 & 1 \\
7 & 2 & 235.1 & 6.13 & 8.77 & 2 & 1 \\
3 & 4 & 215.4 & 5.89 & 8.55 & 2 & 1 \\
3 & 2 & 238.1 & 5.83 & 8.73 & 1 & 1 \\
\bottomrule
\end{tabular}
\caption{Ablation study of distillation setups with varying chunk sizes and diffusion steps based on our internal eval dataset. GPU columns indicate the number of H100 GPUs used for the score model and the VAE decoder respectively.}
\label{tab:ablation}
\end{table}

Overall, we observe that the lip-sync quality (as indicated by Sync-C and Sync-D) remains relatively consistent across different settings, with only minor fluctuations. This suggests that all configurations offer robust lip-sync performance. The perceptual quality, as measured by FVD, shows a slight degradation when both the chunk size is reduced from 7 to 3 and the number of diffusion steps is reduced from 4 to 2. However, this degradation is modest.

What is particularly notable is the trade-off in computational cost. The most compute-efficient setting (chunk size 3 with 2 diffusion steps) requires only 1 H100 GPU for the score model, compared to 4 H100s in the highest quality setting. This results in a significant reduction in hardware cost and energy consumption, while still delivering acceptable generation quality.

Thus, one may choose a configuration based on their specific compute budget and quality expectations. If resource constraints are critical, the 3×2 setup provides a compelling balance between performance and efficiency; if maximum perceptual fidelity is required, the 7×4 setup is preferred.

\section{Applications}

We demonstrate a real-time FaceTime-style application that integrates TalkingMachines with audio large language models (LLMs) to showcase the practical deployment of our system in interactive video communication scenarios.

\subsection{System Architecture}

Our demonstration system consists of three main components:

\begin{itemize}
\item \textbf{Audio LLM Integration:} We integrate mainstream audio LLMs that generate real-time spoken responses, enabling natural conversational interactions with users.

\item \textbf{Video Generation Server:} Our TalkingMachines model is deployed on a cloud server with H100 GPUs, where the score model and VAE decoder operate on separate GPU resources to generate synchronized lip-synced animations from audio input.

\item \textbf{WebRTC Streaming:} We utilize LiveKit~\cite{livekit}, a cloud-based WebRTC service, to handle real-time video streaming and client connections.
\end{itemize}

\subsection{System Workflow}

The system operates through a distributed pipeline where user audio is captured via the web interface, processed by the audio LLM to generate conversational responses, and forwarded to our video generation server. The generated video frames are synchronized with audio and streamed back to clients through the WebRTC service, enabling real-time interactive conversations with AI-generated avatars.

\subsection{Demonstration Results}

Our implementation successfully achieves real-time performance suitable for interactive video calls, with users able to access the system through standard web browsers across desktop and mobile devices. The system demonstrates the practical viability of deploying advanced video generation models in real-time communication applications, establishing a foundation for AI-powered interactive media experiences.

\section{Conclusion, Limitations and Future Work}

In this work, we demonstrate how audio-driven animation and sparse causal distillation can be effectively applied to a powerful pretrained video generation model, transforming it into a streaming-capable network that supports real-time, infinite-length FaceTime-style video generation. The resulting model is capable of animating images of arbitrary styles—including photorealistic, anime, and 3D avatars—with natural, high-quality lip-synced speech, when paired with mainstream audio large language models (LLMs).

We also detail the system-level engineering efforts required to reduce computational bottlenecks in the context of real-time streaming, including optimizations in GPU allocation, communication-computation overlap, and memory reuse. These design choices are critical for enabling the model to run with minimal latency in practical deployment scenarios.

Our training methodology is designed to strike a strong balance between leveraging the generalization capabilities of the pretrained model and extending it with new audio-driven behaviors. Notably, our adaptation introduces significant capabilities—such as expressive lip synchronization and dynamic motion range—using only a modest amount of paired audio-video training data and compute resources.

Despite these advantages, our current approach has limitations. The audio conditioning components were introduced only in the later stages of training, meaning that the pretrained model did not benefit from large-scale audio-video supervision during its initial learning. The large audio projection layers, while effective, were trained on a relatively small subset of data with limited iterations. This bottleneck constrains the model’s scalability and expressiveness in more diverse or challenging audio-driven scenarios.

Looking forward, it would be promising to explore large-scale pretraining strategies that incorporate audio conditions earlier in the process. In particular, joint modeling of video and audio from massive paired datasets may enable even stronger multimodal representations, improved lip-sync fidelity, and more robust performance across domains and languages.

\section*{Project Contributions}

\begin{itemize}
\item \textbf{Project Leader:} Weimin Wang
\item \textbf{Modeling \& Deployment:} Chetwin Low, Weimin Wang
\item \textbf{Data:} Chetwin Low, Weimin Wang, Ryan Vilim, Diego De La Torre, Calder Katyal, Zhiwei Jia, Michal Wolski
\item \textbf{WebRTC(LiveKit):} Yi Cui
\end{itemize}


\bibliographystyle{plain}  
\bibliography{bibliography}
\end{document}